# Measurement of transition frequencies and hyperfine constants of molecular iodine at 520.2 nm

AKIKO NISHIYAMA,[1] SHO OKUBO,[1] TAKUMI KOBAYASHI,[1] AKIO KAWASAKI,[1] AND HAJIME INABA[1]

[1] *National Metrology Institute of Japan (NMIJ), National Institute of Advanced Industrial Science and Technology, 1-1-1 Umezono, 305-8563, Tsukuba, Ibaraki, Japan*
*a-nishiyama@aist.go.jp*

**Abstract:** We measured the transition frequencies of the hyperfine components in the four lines (P(34) 39-0, R(36) 39-0, P(33) 39-0, and R(35) 39-0) of the B-X transitions of molecular iodine at 520.2 nm. The 520.2 nm laser was generated by wavelength-converting the output of a 1560.6 nm external-cavity diode laser using a dual-pitch periodically poled lithium niobate (PPLN) waveguide. The frequencies were measured by counting the heterodyne beats between the laser stabilized at the frequencies of the hyperfine components and a frequency comb synchronized with a hydrogen maser. We determined the transition frequencies of the $a_1$ components with relative uncertainties of $1 \times 10^{-11}$; the uncertainty was limited by the impurity of the molecular iodine in the cell. From the measured hyperfine splitting frequencies, we calculated the hyperfine constants of these four transitions to obtain the rotational dependence of the excited-state hyperfine constants.

## 1. Introduction

Molecular iodine has many rovibronic lines of the B-X transition in the visible wavelength range. Each transition is split into 15 or 21 hyperfine components with narrow linewidths. Therefore, the transition frequencies of molecular iodine have been studied as wavelength standards/references in many fields including metrology [1], and high-resolution molecular spectroscopy [2,3]. In the 500 nm region near the B-state dissociation limit, the effect of predissociation is smaller than in the transitions to lower vibrational states; hyperfine spectra with narrower spectral linewidths [4] can be obtained, allowing the realization of iodine stabilized lasers with high frequency stability [5–7]. The splitting frequencies of the hyperfine structure components in a transition have been measured to determine the hyperfine structure constants of the transition [8,9]. The hyperfine structure constants over a wide wavelength range allow us to determine the dependence of the hyperfine structure interactions on the vibrational and rotational levels in the excited states. This in turn can be used to verify that the measured splitting frequencies are correct. The development of the optical frequency comb has also made it possible to measure precisely the absolute frequencies of the transitions [10–16].

In addition, laser technology is important for high-resolution spectroscopy. For precision spectroscopy, such as the observation of the hyperfine structures of the B-X transitions of molecular iodine, the spectroscopic laser must oscillate at a single frequency and have a narrow spectral linewidth and good frequency controllability. In the past, there were few such lasers, with only helium-neon lasers at 633 nm, Nd-YAG lasers at 532 nm, and argon-ion lasers at 515 nm generally available. Dye lasers and laser diodes (LD) have been widely used, but they require sophisticated handling techniques such as pre-stabilization of the laser frequency or cannot be used at the desired wavelengths. In recent years, significant technological advances have been made on near-infrared LDs and nonlinear optical crystals for harmonic generation. For example, by using the LDs in combination with a periodically poled lithium niobate (PPLN) waveguide, the third harmonic of a frequency-stabilized laser in the telecommunications band can be used to observe the absorption lines of molecular iodine at 514 nm [17]. Thanks to this, the iodine absorption lines are now accessible by using lasers in the telecommunications band, which are inexpensive and have good controllability.

Recently, the demand for frequency standards in the telecommunications wavelength bands has increased, driven by the growing need for high-speed, high-capacity data communications over optical networks. The $v_1 + v_3$ band of $^{13}C_2H_2$ transitions is currently recommended as the frequency standard for the 1.54 um region [18], but it requires cavity enhanced saturation absorption spectroscopy due to the low transition intensity, and the transition band covers only part of the telecommunications band (1520 nm to 1550 nm). In contrast, using the THG of the near-infrared laser source, the rich spectrum of the molecular iodine covers all the wavelengths of the C and L bands (1530 nm to1625 nm) that is used for long-distance fiber transmission. Thanks to its high transition intensity, the signal can be obtained by simple saturated absorption spectroscopy, making the device compact and easily portable. Therefore, iodine-stabilized lasers are expected to become even more important and widely used as simple frequency standards for telecommunication wavelengths. The frequency-stabilized laser at 1560 nm may be particularly attractive for various applications, as it is the center wavelength of mode-locked erbium-doped fiber lasers, which are the most commonly used source as optical frequency combs.

In this study, we developed an iodine stabilized laser using a 1560 nm LD combined with a PPLN waveguide for 520 nm generation. As frequency standards for 520.2 nm and 1560.6 nm, the frequencies of hyperfine structure components in four lines (P(34) 39-0, R(36) 39-0, P(33) 39-0, and R(35) 39-0) of the B-X transition of molecular iodine were determined for the first time using an optical frequency comb. Based on these measurements, we determined the hyperfine constants of the iodine molecules from the frequency difference of the hyperfine splitting of each transition. The results confirm the accuracy of our frequency measurements and provide the rotational dependence of the hyperfine constants of the excited states.

## 2. Experiment

### 2.1 Setup

A schematic diagram of the spectroscopy system is shown in Fig. 1. A narrow linewidth external cavity diode laser (ECDL) (RIO PLANEX) with an output power of about 10 mW and a wavelength of 1560.6 nm was used as the light source. The output of the ECDL was amplified to 420 mW with an Er-doped fiber amplifier and launched into a PPLN waveguide (NTT Innovative Devices Corporation). We obtained 60 mW of third harmonic generation (THG), corresponding to an efficiency of 80 %/W$^2$. The conversion efficiency was calculated as $P_{3\omega}/P_{\omega}^3$, where $P_{3\omega}$ is THG power and $P_{\omega}$ is fundamental power. We developed a spectroscopy system based on the modulation transfer technique [19,20]. The THG output was used; the second harmonic at 780 nm was removed with a dichroic filter. The fundamental light passes through the filter, which does not interfere with the spectroscopy signal. The output beam was split into pump and probe beams for saturated absorption spectroscopy. The pump light was modulated at 220 kHz with an electro-optical modulator (EOM); the probe light was frequency shifted at 80 MHz with an acousto-optic modulator (AOM). The counter-propagating pump and probe beams were incident on the iodine cell with beam diameters of approximately 3 mm. The pump and probe powers in front of the cell were 3 mW and 0.3 mW, respectively. The iodine cell was 20 cm long and the cold finger temperature was stabilized at −10 °C. To extend the interaction length between the molecular iodine and the laser beams and to obtain a high signal-to-noise ratio (SNR) in the observed spectra, we used a three-pass configuration of pump and probe beams [21]. The 520 nm laser frequency was rapidly controlled by changing the injection current into the fundamental 1560 nm ECLD and extensively by controlling the temperature of the ECDL.

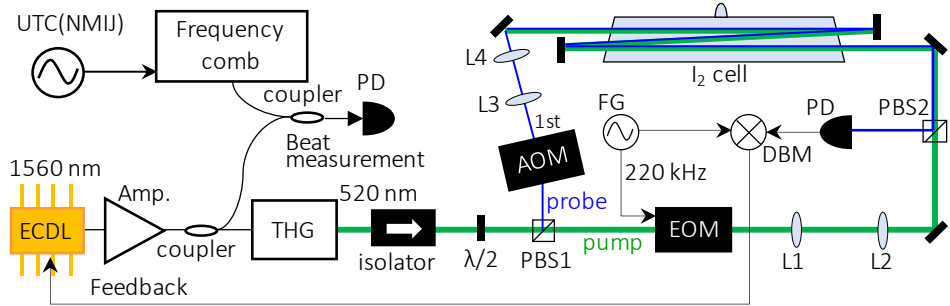

Fig. 1. Schematic diagram of the spectroscopy system. ECDL: external cavity diode laser, Amp.: Er-doped fiber amplifier, THG: third harmonic generation with PPLN waveguide, PBS1, 2: polarization beam splitters, L1-4: lenses, EOM: electro-optical modulator, AOM: acousto-optic modulator, PD: photo detector, DBM double-balanced mixer, FG: function generator.

*2.2 Spectra observation, laser frequency stabilization, and frequency measurement*

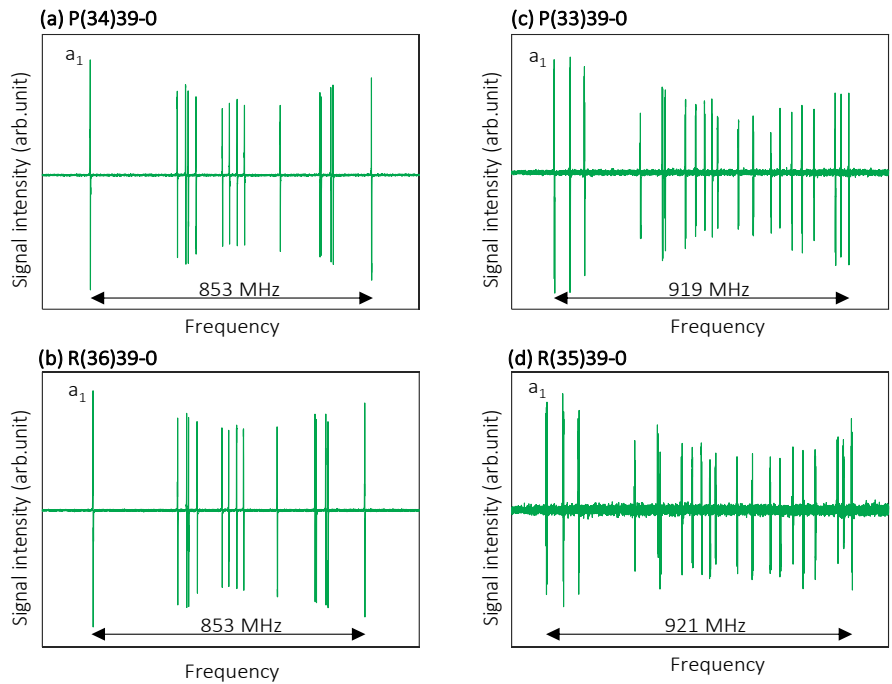

Fig. 2. Observed hyperfine spectra with saturated absorption and modulation transfer spectroscopy. (a) P(34) 39-0, (b) R(36) 39-0, (c) P(33) 39-0, (d) R(35) 39-0.

First, we observed four transitions by slowly sweeping the laser frequency at 520.2 nm. Figure 2 shows the observed spectra of the four transitions: (a) P(34) 39-0, (b) R(36) 39-0, (c) P(33) 39-0, (d) R(35) 39-0. In the transitions we observed 15 hyperfine components with even *J*-numbers and 21 with odd *J*-numbers. All the hyperfine components were well resolved. The observed spectra show dispersion line shapes since the signals were obtained by modulation transfer spectroscopy. The hyperfine spectra in (c) and (d) have more noise than in (a) and (b).

It can be assumed that the noise in the spectra was increased because the intensity noise of the ECDL was higher in a certain wavelength range.

The laser frequency was stabilized at each of these hyperfine components in turn by using the modulation transfer spectroscopy signal as an error signal for feedback control of the laser current. The laser frequency was then measured using a home-made frequency comb based on erbium-doped fibers [22]. For each frequency measurement, we counted the frequency of a beat note at 1560 nm between a portion of the amplified ECDL output and the comb phase stabilized to a hydrogen maser (See Fig. 1). The mode number of the comb interfering with the ECDL was determined using a wavelength meter whose precision (~10 MHz) was sufficiently smaller than the comb's repetition rate (~120 MHz). For confirmation, we measured the ECDL frequency using another comb with a different repetition rate to check for consistency. Each beat frequency was measured with two frequency counters to verify measurement consistency. The repetition rate $f_{\text{rep}}$ was also measured with a single counter, but the set value was used in the optical frequency calculation. This is because the actual $f_{\text{rep}}$ variation was less than the counter's measurement uncertainty. In addition, we also measured the carrier envelope offset frequency with a single counter and used the results in the optical frequency calculation.

## 3. Results and discussion

### 3.1 Frequency stability of laser stabilized to hyperfine components

Figure 3 shows the Allan deviation as the frequency stability of a laser stabilized to the $a_1$ component of P(34) 39-0 calculated from the beat frequency between the stabilized laser and the frequency comb. The stability of a hydrogen maser is also plotted; for averaging times of 1 s to 100 s, the Allan deviation is limited by the stability of the hydrogen maser. By contrast, the stability at averaging times exceeding 100 s is limited by the various long-term variations of the spectroscopy system. The instability at an average time of 100 s is $6 \times 10^{-15}$, which is sufficiently small compared with the measurement uncertainty of our setup. In this study, we choose 120 s as the averaging time when we measured all the transition/split frequencies discussed in Section 3.2.

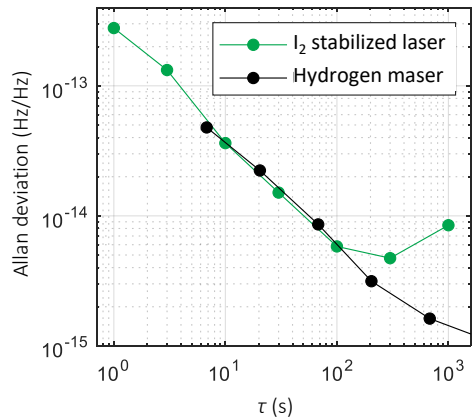

Fig. 3. Allan deviation of the iodine-stabilized laser stabilized to the $a_1$ components of the P(34) 39-0 (green) and the hydrogen maser (black) used as a reference for the optical frequency comb.

### 3.2 Transition frequencies of $a_1$ components

We determined the transition frequencies of the $a_1$ components of the four transitions (Table 1). These are based on optical frequency measurements using a frequency comb; the power and pressure shifts are corrected. These, including uncertainties, are discussed in detail in the next subsection.

Table 1. Transition frequencies of $a_1$ components of P(34) 39-0, R(36) 39-0, P(33) 39-0, and R(35) 39-0

| Transition | Transition frequency (kHz) |
|---|---|
| P(34) 39-0 $a_1$ component | 576 280 542 700.5 (5.1) |
| R(36) 39-0 $a_1$ component | 576 292 529 065.3 (5.2) |
| P(33) 39-0 $a_1$ component | 576 316 835 077.8 (5.3) |
| R(35) 39-0 $a_1$ component | 576 328 505 018.7 (5.5) |

*The pressure and power shifts are corrected.

## 3.3 Power shift, pressure shift, and uncertainty budget of transition frequencies

We investigated several systematic shifts and systematic uncertainties in the determination of the transition frequency of each transition (Table 2).

Table 2. Systematic frequency shifts and uncertainties

| | | P(34) 39-0 | | |
|---|---|---|---|---|
| Effect | Coefficient, etc. | Measurement condition, uncertainty | Correction ($v_{corrected} - v_{measured}$) / kHz | Frequency uncertainty $f_i$ / kHz ($k = 1$) |
| Iodine vapor pressure shift | −5.00(16) kHz/Pa | 1.44(16) Pa | 7.2 | 0.9 |
| Power shift | 0.44(5) kHz/mW | 3.3(2) mW | −1.4 | 0.2 |
| Misalignment of pump and probe beams | 1.7 kHz/mrad | < 0.5 mrad | 0 | < 0.9 |
| Laser frequency servo offset | | < 0.1 kHz | 0 | < 0.1 |
| EOM phase mismatch | 0.19 kHz/rad | < $\pi$/18 rad | 0 | < 0.03 |
| Frequency measurement | < 2.1×10$^{-14}$ | 120 s averaging | 0.3 | < 0.01 |
| Cell impurity | | 5 kHz | 0 | 5 |
| Total | | | 6.0 | 5.1 |
| Relative | | | | 8.9 × 10$^{-12}$ |

| | | R(36) 39-0 | | |
|---|---|---|---|---|
| Iodine vapor pressure shift | −4.80(13) kHz/Pa | 1.44(16) Pa | 6.9 | 0.8 |
| Power shift | 0.51(11) kHz/mW | 3.0(2) mW | −1.7 | 0.4 |
| Misalignment of pump and probe beams | 1.7 kHz/mrad | < 0.5 mrad | 0 | < 0.9 |
| Laser frequency servo offset | | < 0.3 kHz | 0 | < 0.3 |
| EOM phase mismatch | 0.42 kHz/rad | < $\pi$/18 rad | 0 | < 0.07 |
| Frequency measurement | < 2.1×10$^{-14}$ | 120 s averaging | 0.3 | < 0.01 |
| Cell impurity | | 5 kHz | 0 | 5 |
| Total | | | 5.5 | 5.2 |
| Relative | | | | 8.9 × 10$^{-12}$ |

| | | P(33) 39-0 | | |
|---|---|---|---|---|
| Iodine vapor pressure shift | −5.03(47) kHz/Pa | 1.44(16) Pa | 7.2 | 1.1 |
| Power shift | 0.79(12) kHz/mW | 3.0(2) mW | −2.6 | 0.4 |
| Misalignment of pump and probe beams | 1.7 kHz/mrad | < 0.5 mrad | 0 | < 0.9 |
| Laser frequency servo offset | | < 0.5 kHz | 0 | < 0.5 |
| EOM phase mismatch | 4.8 kHz/rad | < $\pi$/18 rad | 0 | < 0.8 |
| Frequency measurement | < $2.1 \times 10^{-14}$ | 120 s averaging | 0.3 | < 0.01 |
| Cell impurity | | 5 kHz | 0 | 5 |
| Total | | | 4.9 | 5.3 |
| Relative | | | | $9.2 \times 10^{-12}$ |
| | | R(35) 39-0 | | |
| Iodine vapor pressure shift | −5.23(54) kHz/Pa | 1.44(16) Pa | 7.5 | 1.2 |
| Power shift | 0.96(20) kHz/mW | 3.0(2) mW | −3.2 | 0.7 |
| Misalignment of pump and probe beams | 1.7 kHz/mrad | < 0.5 mrad | 0 | < 0.9 |
| Laser frequency servo offset | | < 1.1 kHz | 0 | < 1.1 |
| EOM phase mismatch | 6.7 kHz/rad | < $\pi$/18 rad | 0 | < 1.2 |
| Frequency measurement | < $2.1 \times 10^{-14}$ | 120 s averaging | 0.3 | < 0.01 |
| Cell impurity | | 5 kHz | 0 | 5 |
| Total | | | 4.6 | 5.5 |
| Relative | | | | $9.5 \times 10^{-12}$ |

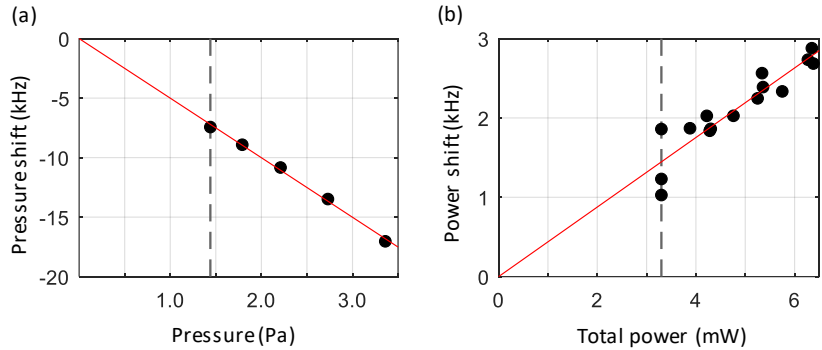

Fig. 4. Center frequency shift of P(34) 39-0 $a_1$ component vs (a) iodine cell pressure and (b) total power. Observed frequencies (black circle) and linear fits (red line).

The transition frequency of molecular iodine varies with the vapor pressure in the cell. We measured the pressure shift coefficient of the $a_1$ component for each transition and also determined the pressure-shifted frequency and its uncertainty. To calculate the coefficients, we measured the laser frequencies while varying the pressure in the iodine cell from 1.44 Pa to 3.36 Pa (corresponding to a cold finger temperature of −10 °C to −2 °C [23]). Figure 4(a) shows the pressure-shifted frequencies of the $a_1$ component of the P(34) 39-0 transition. The measured frequencies are shown as black circles. The pressure shift coefficient obtained by least-squares fitting was −5.00(16) kHz/Pa. The frequency measurements needed to determine the transition frequency of the $a_1$ component were performed at a vapor pressure of 1.44(16) Pa. The uncertainty of the vapor pressure was estimated from the specification of the thermistor used to measure the cold finger temperature. Therefore, the pressure-shifted frequency at a vapor pressure of 1.44 Pa is −7.2 kHz, and the uncertainty is 0.83 kHz [24]. The pressure-shifted frequencies and uncertainties for the other transitions are listed in Table 2.

The transition frequency also varies with the intensity of the laser beam passing through the iodine vapor. We refer to the changed frequency as a power shift. To determine the power shift coefficients, we varied the pump power for molecular iodine from 3 mW to 6 mW and measured the frequency of the stabilized laser. The probe laser power was set at 0.3 mW; the beam diameters of the both the pump and probe lasers in the iodine cell were approximately 3 mm. Figure 4(b) shows the power shift of the $a_1$ component of the P(34) 39-0 transition. The proportional coefficient between the power shift and the laser power obtained by least-squares fitting was 0.44(5) kHz/mW. The optical frequency measurements needed to determine the transition frequency were performed at a total laser power of 3.3(2) mW. The power uncertainties were estimated from the power meter specifications. Therefore, the power shift at a total power of 3.3 mW was 1.5 kHz, and the uncertainty was 0.2 kHz [24]. The power shifts and uncertainties for the other transitions are listed in Table 2.

Other systematic frequency uncertainties were also investigated. Here we describe an example with the P(34) 39-0 transition. The results for the other three transitions are listed in Table 2. The misalignment of the pump and probe beams causes a frequency shift. The measured angular dependence of this shift was 1.7 kHz/mrad. The frequency uncertainty is estimated to be < 0.9 kHz for all the transitions, corresponding to the uncertainty in the pointing direction of the pump beam of < 0.5 mrad. The frequency shift due to the electrical offset of the servo was estimated to be < 0.1 kHz. We measured the frequency shift caused by the demodulation phase adjustment. The phase shifts between the applied modulation and the demodulation caused laser frequency shifts. The uncertainty due to this effect was less than 0.03 kHz. The other transitions had relatively larger uncertainties of electrical offset of the servo and phase adjustment as shown in Table 2. The uncertainty of the frequency reference used for frequency combs is caused by the hydrogen maser, and the relative uncertainty at an average time of 120 s is $2.1 \times 10^{-14}$, which is negligible in this study. The hydrogen maser had an offset of $4.3 \times 10^{-13}$ from UTC, resulting in a shift of 0.25 kHz. Iodine cell contamination causes a frequency shift. This effect can add an uncertainty of 5 kHz to the measurement results, which is dominant in the error budget [25].

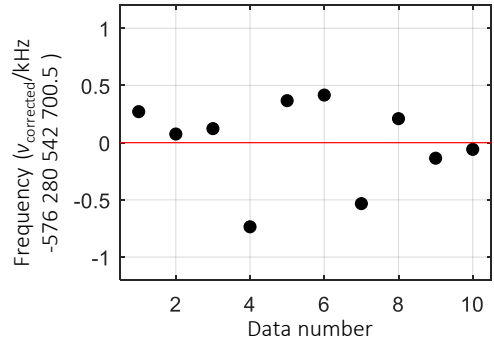

Fig. 5. Measured frequencies of the laser stabilized at the $a_1$ component of the P(34)39-0 transition; the power and pressure shifts are corrected. The zero frequency is the determined transition frequency of the $a_1$ component. The measurements were repeated 10 times on different days.

To verify the above uncertainty budget, we measured the frequency of the laser stabilized to the $a_1$ component of the P(34) 39-0 transition 10 times with readjustments of the optical alignment and electrical offsets. The measurements were performed on different days under the same conditions, namely a cell pressure of 1.44 Pa and a pump power of 3.0 mW. The results are shown in Fig. 5. The standard deviation was 0.38 kHz. The main sources of variation were considered to be the realignment of the pump probe beam and the electrical offset of the servo. The measurement result is consistent with the uncertainty budget.

*3.4 Hyperfine constants and their rotational dependence*

We determined the hyperfine structure constants and the rotational dependence using the transition frequencies of all the components of each transition. The hyperfine Hamiltonian of the iodine molecule is represented by the effective Hamiltonian [26]. The effective Hamiltonian ($H_{hf}$) can be written with four major contributions from the nuclear electric quadrupole ($H_{eQq}$), spin–rotation ($H_{SR}$), tensorial spin–spin ($H_{TSS}$), and scalar spin–spin interactions ($H_{SSS}$) as

$$H_{hf} = eQq\, H_{eQq} + C\, H_{SR} + d\, H_{TSS} + \delta\, H_{SSS}.$$

The observed hyperfine splittings are fitted to the effective Hamiltonian. Table 3 shows the hyperfine splitting (Obs.) and those reproduced from the hyperfine constants obtained by fitting (Calc.) for the four transitions. It also shows the difference between Obs. and Calc.; the standard deviations of the offsets of the fitted values from the observed values are shown at the bottom. For the P(33) 39-0 and R(35) 39-0 transitions, the standard deviations of the fit are somewhat larger. This is probably due to the low SNR of the spectra observed at these transitions (see Figs. 2(c) and 2(d)).

Table 3. Hyperfine splitting of the P(34)39-0, R(36)39-0, R(33)39-0, and R(35)39-0 transitions

| | P(34) 39-0 | | | R(36) 39-0 | | |
|---|---|---|---|---|---|---|
| | Obs. (kHz) | Calc. (kHz) | Diff. (kHz) | Obs. (kHz) | Calc. (kHz) | Diff. (kHz) |
| $a_1$ | 0.0 | -1.3 | 1.3 | 0.0 | -1.6 | 1.6 |
| $a_2$ | 255 294.5 | 255 294.1 | 0.5 | 253 418.7 | 253 417.2 | 1.6 |
| $a_3$ | 280 225.4 | 280 225.5 | -0.1 | 281 145.8 | 281 145.8 | 0.0 |
| $a_4$ | 287 987.5 | 287 989.0 | -1.4 | 287 088.8 | 287 091.0 | -2.2 |
| $a_5$ | 311 678.1 | 311 678.4 | -0.3 | 313 619.4 | 313 619.7 | -0.3 |
| $a_6$ | 392 081.6 | 392 082.6 | -1.0 | 391 178.6 | 391 179.2 | -0.7 |
| $a_7$ | 413 380.7 | 413 382.3 | -1.6 | 412 432.1 | 412 435.0 | -2.9 |
| $a_8$ | 436 566.1 | 436 568.0 | -1.8 | 437 718.0 | 437 718.5 | -0.5 |
| $a_9$ | 458 697.2 | 458 697.0 | 0.2 | 459 703.4 | 459 704.6 | -1.3 |
| $a_{10}$ | 567 674.4 | 567 674.2 | 0.2 | 567 745.0 | 567 744.7 | 0.3 |
| $a_{11}$ | 690 488.2 | 690 488.3 | -0.1 | 689 774.3 | 689 776.5 | -2.2 |
| $a_{12}$ | 694 175.3 | 694 174.7 | 0.5 | 694 112.6 | 694 111.3 | 1.3 |
| $a_{13}$ | 724 901.9 | 724 900.9 | 1.0 | 725 142.9 | 725 141.8 | 1.2 |
| $a_{14}$ | 731 941.6 | 731 942.4 | -0.9 | 732 622.6 | 732 623.5 | -0.9 |
| $a_{15}$ | 852 811.5 | 852 811.5 | 0.0 | 852 857.6 | 852 856.3 | 1.3 |
| | Standard deviation of the Diffs: 1.1 kHz | | | Standard deviation of the Diffs: 1.8 kHz | | |
| | P(33) 39-0 | | | R(35) 39-0 | | |
| | Obs. (kHz) | Calc. (kHz) | Diff. (kHz) | Obs. (kHz) | Calc. (kHz) | Diff. (kHz) |
| $a_1$ | 0.0 | -4.0 | 4.0 | 0.0 | -4.9 | 4.9 |
| $a_2$ | 46 471.0 | 46 467.0 | 4.1 | 48 378.7 | 48 373.1 | 5.6 |
| $a_3$ | 89 323.4 | 89 322.0 | 1.4 | 93 256.5 | 93 253.4 | 3.2 |
| $a_4$ | 258 209.8 | 258 209.9 | 0.0 | 260 051.3 | 260 052.1 | -0.8 |
| $a_5$ | 326 288.4 | 326 287.6 | 0.8 | 329 138.5 | 329 135.5 | 2.9 |
| $a_6$ | 334 834.7 | 334 835.2 | -0.5 | 335 824.9 | 335 828.7 | -3.8 |
| $a_7$ | 398 407.2 | 398 408.9 | -1.7 | 400 611.2 | 400 614.4 | -3.2 |
| $a_8$ | 430 611.7 | 430 615.0 | -3.2 | 432 327.1 | 432 331.4 | -4.3 |
| $a_9$ | 459 724.8 | 459 727.8 | -3.0 | 460 738.9 | 460 741.4 | -2.4 |
| $a_{10}$ | 482 867.7 | 482 868.7 | -1.0 | 485 912.1 | 485 915.4 | -3.3 |
| $a_{11}$ | 500 382.4 | 500 385.6 | -3.1 | 503 598.3 | 503 602.6 | -4.3 |
| $a_{12}$ | 566 615.7 | 566 618.1 | -2.3 | 567 963.4 | 567 966.2 | -2.8 |
| $a_{13}$ | 614 060.9 | 614 062.3 | -1.4 | 616 050.1 | 616 052.0 | -1.9 |
| $a_{14}$ | 670 383.5 | 670 384.5 | -1.0 | 672 245.8 | 672 247.6 | -1.8 |
| $a_{15}$ | 698 984.8 | 698 986.6 | -1.8 | 701 814.3 | 701 816.5 | -2.2 |
| $a_{16}$ | 737 278.4 | 737 277.6 | 0.8 | 739 441.3 | 739 440.2 | 1.1 |
| $a_{17}$ | 769 681.2 | 769 679.6 | 1.7 | 771 945.1 | 771 942.2 | 2.9 |
| $a_{18}$ | 807 612.0 | 807 610.0 | 2.0 | 809 114.3 | 809 111.8 | 2.5 |
| $a_{19}$ | 875 917.2 | 875 913.3 | 4.0 | 878 039.2 | 878 033.8 | 5.3 |
| $a_{20}$ | 893 548.8 | 893 548.8 | 0.0 | 895 882.4 | 895 880.8 | 1.6 |
| $a_{21}$ | 918 954.6 | 918 949.6 | 5.0 | 920 937.8 | 920 933.0 | 4.8 |
| | Standard deviation of the Diffs: 2.8 kHz | | | Standard deviation of the Diffs: 3.9 kHz | | |

Since the effective hyperfine Hamiltonian can reproduce the experimental values with high accuracy, an evaluation of the difference between the fitted hyperfine splitting and the observed values provides a guarantee of the measurement accuracy. The hyperfine constants, which are the coefficients of each term of the hyperfine Hamiltonians, were fitted to the observed hyperfine spectra. The fitting program was written based on refs. [27-29]. As shown in Table 3, the standard deviations of the fit in this study are comparable to those of previous studies of other transitions [13,30,31], suggesting that the frequency measurement results are reasonable.

**Table 4. Fitted hyperfine constants**

| | $\Delta eQq$ (MHz) | $\Delta C$ (kHz) | $\Delta d$ (kHz) | $\Delta \delta$ (kHz) |
|---|---|---|---|---|
| P(34) 39-0 | 1898.753(3) | 138.881(4) | -73.7(1) | -9.1(2) |
| R(36) 39-0 | 1898.719(4) | 139.232(6) | -74.0(2) | -9.0(3) |
| P(33) 39-0 | 1898.750(5) | 138.809(7) | -73.7(3) | -8.5(2) |
| R(35) 39-0 | 1898.722(6) | 139.143(9) | -74.1(4) | -8.4(3) |

Table 4 shows the values of the four hyperfine constants of the four transitions. The fit to the observed spectra gives the difference between the hyperfine constants of the excited and ground states: $\Delta eQq$, $\Delta C$, $\Delta d$, and $\Delta \delta$. We can derive the excited state hyperfine constants using the ground state hyperfine constants from ref. [32]. The resulting rotational dependence of the hyperfine constants of the excited state is plotted against $J'(J'+1)$ in Fig. 6, where $J'$ is the rotational quantum number of the excited state of the transitions. The error bars represent the fitting uncertainty for each constant. The fitting errors of $eqQ$ and $C$ are relatively small and indicate linear dependences on $J'(J'+1)$. In addition, $d$ indicates a slight downward slope. The variation of $\delta$ with $J'(J'+1)$ is small compared with the error bars. Extensive and numerous vibrational rotational dependencies of hyperfine constants in the B state have been reported in ref. [13]. Although the rotational dependence of the hyperfine constants of $v'= 39$ state has not been reported in previous studies, the obtained hyperfine constants are consistent with those expected from the hyperfine constants of other vibrational states [13]. From the fact that the hyperfine splitting is well represented by the effective Hamiltonian and the obtained hyperfine constants show monotonic rotational state dependences, we conclude that there are no specific interactions from other excited states at $v'=39$.

## 4. Summary

In this study, we determined the transition frequencies of the $a_1$ components of four transitions P(34) 39-0, R(36) 39-0, P(33) 39-0, R(35) 39-0 at 520.2 nm using the THG of a narrow linewidth ECDL at 1560.6 nm. The systematic shift and uncertainties are evaluated, and the relative uncertainties of transition frequency determination was $1 \times 10^{-11}$. The pressure and power shift measurements increase the usefulness of the measured transitions as frequency standards. We analyzed the hyperfine splitting of the four transitions and derived the hyperfine constants with small fitting errors. The results also successfully showed the rotational dependence of the hyperfine constants of the excited state. The transition frequencies determined in this study can be used as frequency standards at 520.2 nm and in the telecommunication wavelength band of 1560.6 nm. The results of this study will contribute to work on wavelength standards and to the fundamental science of the hyperfine interaction of molecular iodine.

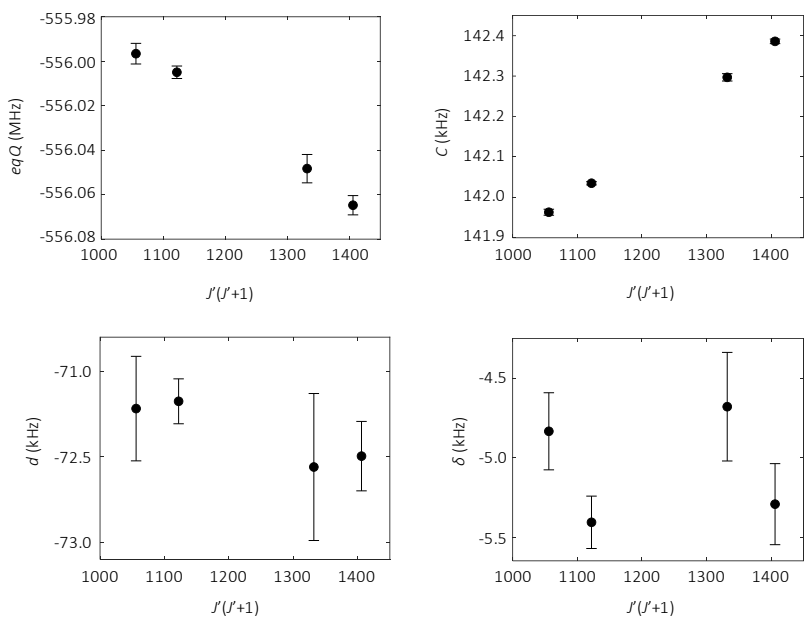

Fig. 6. Rotational dependence of the hyperfine constants of excited state.

**Funding.** Research Foundation for Opto-Science and Technology; 34th Matsuo Foundation; Moonshot Research and Development Program (JPMJMS2268); Center of Innovation NEXT (JPMJPF2015)

**Acknowledgments.** We are grateful to Dr. Y. Nishida of NTT Innovative Devices Corporation for the design of and helpful discussion as regards the PPLN waveguide. We are also grateful to our colleagues Dr. Tanabe and Dr. Yasuda for helpful discussions as regards the time standards and for providing the hydrogen maser signal.

**Disclosures**
The authors declare no conflicts of interest.